\documentclass[useAMS,usenatbib]{mn2e}
\usepackage{psfig, epsf, epsfig}
\title[Kinematics of the LMC]{On the Origin of the Kinematical
Differences Between the Stellar Halo and the Old Globular Cluster
System in the Large Magellanic Cloud}
\author[K. Bekki]
{Kenji Bekki${}^1$\thanks{E-mail:
bekki@phys.unsw.edu.au} \\
       ${}^1$School of Physics, University of New South Wales,
              Sydney 2052, NSW, Australia}
  
\begin{document}

\date{Accepted, Received 2005 February 20; in original form }

\pagerange{\pageref{firstpage}--\pageref{lastpage}} \pubyear{2005}

\maketitle

\label{firstpage}

\begin{abstract}

We discuss structural and  kinematical properties of
the stellar halo and the old globular cluster system (GCS) in
the Large Magellanic Cloud (LMC) based on numerical simulations
of the LMC formation.
We particularly discuss 
the observed possible GCS's rotational kinematics
($V/\sigma \sim 2$)  that
appears to be  significantly different from the stellar
halo's one with a large velocity dispersion ($\sim 50$ km s$^{-1}$). 
We consider that both halo field stars and old GCs
can originate from  low-mass subhalos virialized at high redshifts ($z >6$).
We investigate the final dynamical properties of the
two old components in the LMC's halo  formed from merging
of low-mass subhalos with field stars and GCs.
We find that the GCS composed of old 
globular clusters (GCs) 
formed at  high redshifts ($z > 6$)
has little rotation ($V/\sigma \sim 0.4$) and structure
and kinematics similar to those of  the stellar halo.
This inconsistency between
the simulated GCS's kinematics 
and the observed one is found to be seen in models with
different parameters. 
This inconsistency therefore
implies that if  old, metal-poor  GCs in  the LMC 
have rotational kinematics, 
they are highly unlikely to originate from 
the low-mass subhalos
that formed the stellar halo.
We thus discuss a scenario in which the stellar halo was formed
from low-mass subhalos
with no/few  GCs whereas the GCS
was formed at the very early epoch of the LMC's disk formation via
dissipative minor and major  merging of gas-rich subhalos  and gas infall.
We also discuss whether old GCs in the LMC
can be slightly younger than the Galactic counterparts.
We suggest that there can be a threshold subhalo mass above 
which  GCs can be formed within subhalos at high redshifts
and thus that this threshold causes differences in physical
properties between stellar halos and GCSs in less luminous galaxies
like the LMC.

\end{abstract}


\section{Introduction}

Structural,  kinematical, chemical  properties of the 
Large Magellanic Cloud (LMC) have been investigated
by many authors concerning different stellar populations
 and gaseous components
and suggested to have valuable information 
on formation and evolution history of the LMC
(Hartwick \& Cowley 1988; Meatheringham et al. 1988;
Irwin 1991; Luks \& Rohlfs 1992; Kunkel et al. 1997;
Graff et al. 2000; Olsen \& Salyk 2002; 
van den Marel et al. 2002; Cioni \& Habing 2003; 
Staveley-smith et al. 2003; Cole et al. 2005).
Previous  numerical studies  on dynamical and chemical
evolution of the LMC compared the results
with these observations and thereby tried  
to provide reasonable physical  explanations for the observations
(e.g., Bekki et al. 2004; Bekki \& Chiba 2005).
Since these studies focused mainly on the origin of disk components
of the LMC, the origin of the old stellar halo and the globular
cluster system (GCS) composed of old, metal-poor globular clusters
(GCs) remains unclear.

Recent observations have reported that
the stellar halo of the LMC has a projected radial density
profile similar to an exponential one (e.g., Alves 2004)
and a larger velocity dispersion 
with possibly little rotation (Minniti et al. 2003; Gratton et al 2004).
The dynamically hot nature of the LMC's stellar halo 
has been confirmed  by
Borissova et al. (2006), which
found that  the stellar halo composed
of RR Lyrae stars has the mean velocity dispersion
of $\sim 50 \pm 2$ km s$^{-1}$ and 
a Gaussian metallicity distributions function with
mean [Fe/H] = $-1.53 \pm 0.02$ dex.
Subramaniam (2006) 
investigated spatial distributions
of RR Lyrae stars from the catalogue by Soszynski et al. (2003) 
and found that the inner stellar halo
can have a disky density distribution.

\begin{table*}
\centering
\begin{minipage}{185mm}
\caption{Model parameters and a brief summary of the results}
\begin{tabular}{cccccccc}
model &
{$M_{\rm t}$ ($\times 10^{10} {\rm M}_{\odot}$) 
\footnote{Initial total masses.}}
& {$\lambda$
\footnote{Initial spin parameters.}}
& {${\delta}_{\rm i}$
\footnote{Initial over densities.}}
& {${\rm N}_{\rm min}$
\footnote{Minimum mass for halo identification.}}
& {$z_{\rm trun}$ 
\footnote{The epoch of truncation of GC formation.}}
& {${(\frac{V}{\sigma})}_{\rm FS}$ 
\footnote{Final $\frac{V}{\sigma}$ for field stars (FS), where
$V$ and ${\sigma}$ are the maximum rotational velocity and
the central velocity dispersion, respectively in the simulated LMC halo.
The man value averaged for three projections ($x$-$y$, $x$-$z$, and $y$-$z$)
is shown.}}
& {${(\frac{V}{\sigma})}_{\rm GC}$  
\footnote{$\frac{V}{\sigma}$ for GCs. }}
\\
Standard  & 6.0 &  0.08  & 0.39 &  32 & 15  & 0.34 & 0.39 \\
Low density  & 6.0 & 0.08  & 0.19 &  32 & 15 & 0.60 & 0.56  \\
High threshold & 6.0   & 0.08 & 0.39 & 1000 & 15 & 0.28  & 0.22  \\
Low-z truncation  & 6.0  & 0.08 & 0.39 & 32 & 10 & 0.19 & 0.21  \\
\end{tabular}
\end{minipage}
\end{table*}

Previous observational studies on the kinematics of the  GCS
composed of old, metal-poor GCs in the LMC 
suggested that the GCS has a disky distribution and rotational
kinematics with a small velocity dispersion (e.g., Freeman et al. 1983;
Kinman et al. 1991;  Schommer et al. 1992; 	
Grocholski et al. 2006).
It is suggested that
it could be difficult to make a robust conclusion
on the presence of the rotational kinematics owing to
the small number (13) of the GCs (van den Bergh 2004).
The total number of old, metal-poor GCs with [Fe/H] $<$ $-1.3$
and well determined radial velocities are five in
the sample of GCs by Grocholski et al. (2006),
which implies that old GCs can have different
kinematics from young and intermediate-age GCs with
rotational kinematics. 
The observed possible rotational kinematics
with $V/\sigma \sim 2$ (Schommer et al. 1992) 
however suggests  that there is a significant kinematical
difference in the stellar halo and the GCS composed
of old GCs in the LMC.
No theoretical explanations however have been proposed for 
the origin of this kinematical difference.

The purpose of this paper is to investigate
the origin of the observed kinematical difference between 
the stellar halo and the GCS composed of old, metal-poor GCs
in the LMC based on numerical simulations of the LMC formation.
We here  consider that the stellar halo and the GCS are formed
from hierarchical merging of subhalos
that are virialized before reionization and have
both field stars and GCs. 
We therefore
perform numerical simulations for the formation of the LMC
based on the cold dark matter (CDM) model 
and thereby investigate structure and kinematics of the simulated
stellar halo and GCS.
By comparing the simulated kinematical properties
of the stellar halo and the GCS with observations,
we try to discuss the relationship between the  stellar  halo formation
and GC one in the early history of the LMC.

 The plan of the paper is as follows: In the next section,
we describe our  numerical models for the stellar halo and GC  formation
of the LMC.
In \S 3, we
present the numerical results
mainly on the kinematics of the simulated halo and GCS
for variously different models.
In \S 4, we discuss  a promising scenario 
explaining  the observe kinematical difference between
the old stellar halo and the GCS in the LMC  
We summarize our  conclusions in \S 5.

\section{The model}

\subsection{GC formation}

This paper is the first step toward better understanding
structural, kinematical, and chemical properties
of the  LMC's GCS
in a comprehensive way.
Therefore we adopt a more idealized model 
of GC formation in the LMC that is assumed to be
formed from hierarchical merging of low-mass subhalos.
We adopt a model in which GCs (and field stars) are formed
within subhalos virialized at high redshifts ($z>6$).
Although the adopted model of GC formation within subhalos at high redshifts
has not been confirmed  
observationally (Brodie \& Strader 2006 for different models of GC formation),
recent numerical simulations based on the model
have successfully explained some fundamental observations
such as radial density profiles of GCSs
(e.g., Santos 2003; Bekki 2005).

We consider that star formation  and thus GC one can proceed
at high redshifts  only in virialized dark matter halos before
reionization.
The physical reason for the suppression of star formation  by
reionization is that ultraviolent background radiation in a reionized
universe can significantly reduce the total amount of cold HI gas
and molecular one (through photoevaporation/photoionization of the gas)
that are observed to be indispensable for active star formation
in low-mass galaxies (e.g., Young and Lo 1997). Recent high-resolution
simulations on this issue (Susa \& Umemura, 2004)
have confirmed that significant suppression of the formation of
cold gas can lead to the suppression of star formation in dwarf galaxies
embedded in dark matter halos,
in particular, lower-mass dwarfs. 
We here focus exclusively on very  old, metal-poor GCs
in the LMC, which could have formed at very high redshifts:
it is, however, observationally unclear whether the GCs were 
formed before reionization.
Therefore, it is reasonable to adopt the above  assumption that
GC formation can proceed in dark matter subhalos virialized before
reionization.

Massive and dense star clusters like GCs  are suggested
to form in GMCs surrounded by high-pressure interstellar medium (ISM)
in galaxies (Elmegreen \& Efremov 1997).
We therefore assume that GCs are formed in the very central regions
of subhalos, where gaseous pressure should be so high 
owing to  their  deep gravitational  potentials.
During destruction of subhalos with GCs in the hierarchical 
growth of the LMC,
the GCs are stripped and
dispersed into the halo to become the halo GCs in the LMC.
This formation processes of GCs from low-mass galactic systems
was demonstrated by previous numerical simulations
(e.g., Bekki \& Freeman 2003; Mizutani et al. 2003; Bekki \& Chiba 2004),
which suggests that the adopted assumption is quite reasonable.

We perform purely dissipationless simulations on galaxy-scale halo
formation via hierarchical merging of subhalos with field stars and GCs.
In these dissipationless simulations, we first identify possible
formation sites of field stars and GCs in low-mass halos at high $z$
and then follow their evolution during hierarchical merging
of the halos till $z=0$.
The final structural and kinematical properties of the stellar
halo and GCS in the simulated LMC are thus determined by
the details of merging histories of subhalos with field stars and
GCs. We consider that as long as the formation sites of
field stars and GCs in low-mass halos are properly modeled,
the present dissipationless models allow us to derive
physical properties of the stellar halo and the GCS in the LMC
in a reasonable way.
The present simulations are different from our previous
chemodynamical  ones (Bekki \& Chiba 2000; 2001)
which investigated both dynamical and chemical evolution
of forming galaxies in order to reproduce structures, kinematics,
and chemical properties  {\it  for  the Galactic field stars}
in a self-consistent manner.
We consider that it is currently difficult
and numerically costly
to construct the fully self-consistent chemodynamical models
in which formation sites {\it both for field stars and GCs}
can be {\it directly} derived for gaseous regions of low-mass halos.
Accordingly, the present study is the first step for better modeling
star and GC formation in the LMC.
Fully self-consistent chemodynamical simulations with
reasonable and realistic
models for star and GC formation from GMCs will be done
in our future works.

\subsection{Identification of field star and GC particles
in hierarchical galaxy formation}

 We simulate the formation of galaxy-scale  halos
in a $\Lambda$CDM Universe with ${\Omega} =0.3$, 
$\Lambda=0.7$, $H_{0}=70$ km $\rm s^{-1}$ ${\rm Mpc}^{-1}$,
and ${\sigma}_{8}=0.9$,
and thereby investigate merging/accretion
histories of subhalos that can contain low mass dwarfs.
The way to set up initial conditions for the numerical simulations
is essentially
the same as that adopted by Katz \& Gunn (1991) and Steinmetz \& M\"uller
(1995). 
We consider an isolated homogeneous, rigidly rotating sphere, on which
small-scale fluctuations according to a CDM power spectrum are superimposed.
The initial total mass ($M_{\rm t}$), radius,
initial overdensity (${\delta}_{i}$), and  spin parameter ($\lambda$)
are set to be free parameters.

Although we investigate variously different  models with 
$6 \times 10^9 {\rm M}_{\odot} \le M_{\rm t} \le
6.0 \times 10^{11} {\rm M}_{\odot}$,
$\lambda = 0.08$, 0.12, and 0.16,
and  ${\delta}_{i}=0.19$ and 0.39,
we mainly show the results of  the ``LMC'' models with
$M_{\rm t}=6.0\times10^{10} {\rm M}_{\odot}$
and $\lambda = 0.08$.
The choice of  $\lambda = 0.08$ 
are  demonstrated to be quite reasonable 
for late-type disk galaxies 
in previous CDM simulations
(e.g., Katz \& Gunn 1991;  Steinmetz \& M\"uller 1995;
Bekki \& Chiba 2000, 2001), and accordingly
we consider that models with $\lambda = 0.08$
are also reasonable for  the present simulations 
for less luminous disk systems like the LMC.
${\delta}_{i}=0.39$
is chosen such that the final central velocity dispersions 
of ``stellar'' components are similar to the observed ones.

The low-mass ($M_{\rm t}=6 \times 10^9 {\rm M}_{\odot}$)
models  show stellar halos with velocity dispersions
being significantly lower than the observed one.
The high-mass ($M_{\rm t}=6 \times 10^{11} {\rm M}_{\odot}$)
models are investigated in order that we can compare the results of
the LMC model with those of ``the Galaxy'' one and thereby
discuss the differences in dynamical properties between the LMC's
GCS and the Galactic GCS.
These results will be discussed in a wider context of stellar halo
and GCS formation
in galaxies with different Hubble types (Bekki \& Chiba 2007, in preparation). 
The details of parameter values in the simulations
including those related to simulation methods
(e.g., softening lengths) are described later.

We start the collisionless simulation at $z_{\rm start}$ (=30) and follow it 
till $z_{\rm end}$ (=1) to identify virialized subhalos
with the densities larger than $170 {\rho}_{\rm c}(z)$,
where ${\rho}_{\rm c}(z)$ is the critical density of the universe, 
at a redshift $z$.
This $170 {\rho}_{\rm c}(z)$ corresponds to the mean mass density
of a collapsed and thus gravitationally bound
object at $z$ (e.g., Padmanabhan 1993).
The minimum number of particles within a virialized subhalo
($N_{\rm min}$) is set to be 32 corresponding to the mass resolution
of 3.8 $\times$ $10^{6}$ $M_{\odot}$ 
for the LMC  models.
This number of 32 is chosen
so that we can find a virialized object at a given $z$ in a robust manner.
The mass resolution of $1.2 \times 10^{5}$ $M_{\odot}$
is chosen such that the masses of GC particles in the simulations
can be consistent with the observed
typical GC mass ($\sim  10^{5}$ $M_{\odot}$) in the Galaxy.

For each individual virialized subhalo
with the virialized redshift of $z_{\rm vir}$,
we estimate a radius ($r_{\rm b}$) within which 20 \%  of the total mass
is included,  and then the particles within $r_{\rm b}$ are labeled 
as ``baryonic'' particles. This procedure for defining baryonic particles
is based on the assumption that energy dissipation via radiative cooling
allows baryon to fall into the deepest potential well of dark-matter halos.
Such baryonic particles in a subhalo will be regarded as candidate
``stellar'' particles to form stellar halos
and GCSs  in the later dynamical stage,
if the subhalo is later destroyed and baryonic particles initially within
the subhalo is dispersed into the galactic halo region.
Thus, the present dissipationless models track the formation of stellar halos
and GCSs
via hierarchical merging of subhalos, although the models are not adequate
to the study of star formation histories in subhalos (as was done
in our previous studies, e.g., Bekki \& Chiba 2001).

 Stellar particles within $r_{\rm b}$ in a subhalo are divided into 
field star (``FS'') particles and ``GC'' ones accordingly to
their locations with respect to the center of the subhalo.
The stellar particle in the very center of a subhalo is identified as 
GC particle whereas stellar particles other than the GC particle
within $r_{\rm b}$ are identified as FS ones.
The initial distributions of FS particles are thus more diffuse than
those of the GC particles  in virialized subhalos.
Massive, compact star clusters like GCs are suggested to be formed
in extraordinary  high-pressure regions, such as the centers of low-mass
galaxies (e.g., Elmegreen 2004). 
Unbound or weakly bound star clusters, which
can evolve into field stars after their disintegration,
can be formed in outer regions of galaxies where gas density
and pressure are low (e.g., Elmegreen 2004).
We thus consider that the adopted model for the distributions
of the two
old stellar components is reasonable.
The differences in initial positions 
with respect to the centers
of subhalos between FS and GC particles can cause
differences in the final distribution with respect
to a galaxy-scale halo formed from the subhalos
between these particles.

In the present model,  each subhalo is assumed to have
only one GC particle that is located in the center
of the halo. One of the main reasons for this is that
GCs in the present simulations are considered to be
formed in nuclear regions of low-mass halos
(i.e., GCs are initially 
either nuclear star clusters or stellar galactic nuclei)
: this assumption
is similar to the scenario proposed by Zinnecker  et al. (1988).
Furthermore,  our analytical arguments suggest
that low-mass dark halos with the masses
of $2 \times 10^8 {\rm M}_{\odot}$,  the luminous masses being 10\% of
the halos,  and specific frequencies of GCs being 5 
can have one GC.  Given that the halos identified as
being virialized before reionization at each time step in the simulations
mostly have masses less that $2 \times 10^8 {\rm M}_{\odot}$,
the above assumption of one GC in a halo can be reasonable.

If the initial number of GC particles 
in each halo is increased,  the initial difference
in spatial distributions between halo and GC particles
becomes less remarkable in the halo: the final 
differences in structural and kinematical properties
between the stellar halo and the GCS in the LMC
therefore becomes less remarkable. Thus the adopted
models with each subhalo having only one nuclear GC
particle is regarded as those showing maximum possible
differences in final dynamical properties between
the stellar halo and the GCS in the present
LMC model.

\subsection{Truncation of GC formation}

Previous theoretical studies have demonstrated that 
ultraviolet background radiation in a reionized
universe can significantly reduce the total amount of cold HI 
and molecular gas 
that are observed to be indispensable for galactic active star formation
(e.g., Susa \& Umemura 2004).
In order to investigate this suppression effects of star and GC formation
on the final structural and kinematics
properties of the simulated stellar halos and GCSs,
we adopt the following idealized assumption:
{\it If a subhalo is virialized after the completion
of the reionization ($z_{\rm reion}$),
star and GC formation is totally suppressed in such a subhalo.}
Then, hypothetical baryonic/stellar  particles in the
subhalos with $z_{\rm vir}$ $<$ $z_{\rm reion}$ will {\it not} be
identified as FS or GC particles in the later stage, but
those in the subhalos with $z_{\rm vir}$ $\ge$ $z_{\rm reion}$
will be regarded as progenitors of visible stellar halos.

Recent WMAP ({\it Wilkinson Microwave Anisotropy Probe})
observations have shown that plausible $z_{\rm reion}$  
ranges from 11 to 30 (Spergel et al. 2003; Kogut et al. 2003)
whereas quasar absorption line studies give the lower limit
of 6.4 for $z_{\rm reion}$ (Fan et al. 2003).
Guided by these observations, we investigate the models
with $z_{\rm reion}$  = 0 (no reionization), 6,  10,  15, and 20.
The  adopted picture of single  epoch of reionization might well be
somewhat oversimplified and less realistic,
however, this idealized model can help  us to elucidate some essential
ingredients of the reionization effects on stellar halo and GC  formation. 
For convenience, the epoch of the truncation of GC formation
by reionization is denoted as $z_{\rm trun}$
in the following.

We  show the results of the models with $z_{\rm trun}=10$ and 15,
firstly because these models  explain
structural properties of the Galactic stellar halo and GCS
(Bekki 2005; Bekki \& Chiba 2005)
and secondly because models with higher  $z_{\rm trun}$
can also better explain the observed properties of GCSs
in early-type galaxies (Bekki et al. 2007). 
Furthermore, the adopted higher $z_{\rm trun}$ are
roughly consistent with
the latest observation
by WMAP suggesting 
that the epoch of reionization 
is $z={10.9}_{-2.3}^{+2.7}$ (Page et al. 2006).
The models with $z_{\rm trun} < 10$ are found to yield
too many GCs ($>1000$) in the present models so that
they can not be consistent with the observed number of
old GCs (13) in the LMC.
Therefore, we mainly discuss the results of the models
with $z_{\rm trun}=15$ in the present study.

It should be stressed here that without truncations of
GC formation after reionization, the present models produce
too many GCs owing to a large number of low-mass halos
virialized after reionization so that they can not be 
consistent with observations. This however does not
necessarily mean that the truncation of GC formation
by reionization is a crucial in better understanding 
the origin of the LMC's GCS: later, selective dynamical destruction
of low-density GCs originating from low-density halos
virialized {\it after reionization} might well
dramatically reduce the overproduced GCs in a model
with no truncation of GC formation by reionization
so that the simulated GC number can be consistent
with the observed one even in such a model.
Such later destruction of GCs is not modeled at all
in the present study.

\begin{figure}
\psfig{file=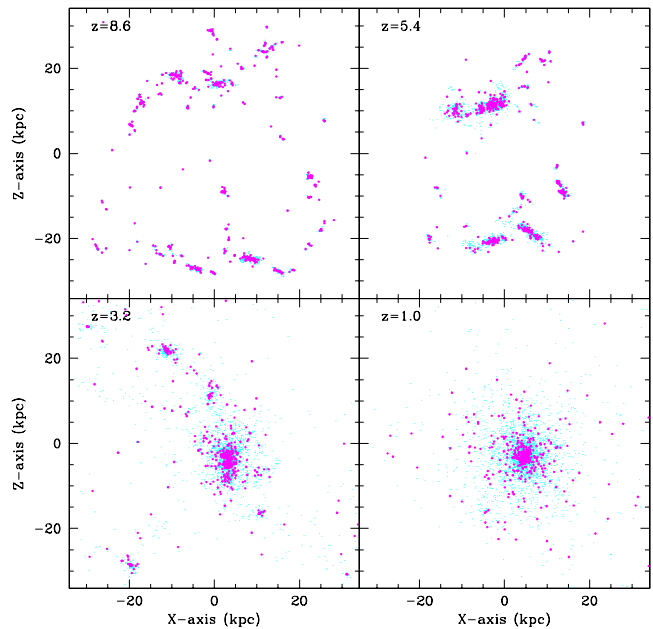,width=8.5cm}
\caption{ 
Time evolution of spatial distributions
of ``FS'' (field star) particles (cyan)
and ``GC'' (globular cluster) ones (magenta)
projected onto the $x$-$z$ plane in
the standard LMC model. For clarity,
GC particles are plotted by bigger dots.
The redshift ($z$) is given in the upper left 
corner for each panel.
}
\label{Figure. 1}
\end{figure}

\subsection{Main points of analysis}

We investigate final structural and kinematical
properties of stellar halos and GCs 
at $z=1$ in models with different model parameters.
We note that later accretion of satellites at $z < 1.5$ is minor
in the final structures of the simulated halos at $z = 0$, so
the calculation is ended at $z_{\rm end} = 1$ to obtain the dynamically
relaxed halo structures.
We first show the ``standard'' LMC  model with
$M_{\rm t}=6.0 \times 10^{10} {\rm M}_{\odot}$,
${\delta}_{\rm i}=0.39$, $\lambda =0.08$, and $z_{\rm trun}=15$ 
then discuss parameter dependences of the results.

We adopt the initial LMC mass significantly
larger than that
of $2 \times 10^{10} {\rm M}_{\odot}$ 
used in the latest simulations for 
the orbital evolution of  {\it the present}  LMC
interacting with the SMC  (e.g., Yoshizawa \& Noguchi 2003).
The reason for this is that numerical simulations 
demonstrated that
the  mass of the LMC can become  significantly
smaller than its initial mass
owing to stripping of dark matter halo and stars
by the Galactic tidal field (Bekki \& Chiba 2005):
We need to adopt the LMC's mass significantly larger
than the present LMC mass. 
The standard model shows that (1) the final  mass
within 7.5 kpc is about $2 \times 10^{10} {\rm M}_{\odot}$
and (2) the central velocity dispersion of the stellar halo
is $50-60$ km s$^{-1}$ depending on projections.
Models with smaller initial masses ($\sim 10^{10} {\rm M}_{\odot}$)
shows that final velocity dispersions of stellar halos 
are significantly smaller than
the observed one.

Although the derived kinematics in the standard model is  
roughly consistent with
observations (e.g., Minniti et al. 2003),
we investigate the following three models for comparison:
The ``low density'' model with ${\delta}_{\rm i}=0.19$,
the ``high threshold'' one with $N_{\rm min}=1000$,
and the ``low-z truncation''  one with $z_{\rm trun}=10$.
The velocity dispersion of the stellar halo 
in the low density model ($30-40$ km s$^{-1}$) 
is significantly smaller than the observed one,
because the final single halo has a lower mean density and thus
have a smaller mass at $z=1$ within 20 kpc that is more reasonable
for the tidal radius in the  more massive LMC
at $z=1$ than the present tidal radius of 
15 kpc (van der Marel et al. 2002).
Only subhalos with masses larger than $10^8 {\rm M}_{\odot}$
at the epochs of virialization can have FSs and GCs 
in the high threshold model.

By assuming that the projected radial density profiles of GCSs
(${\Sigma}_{\rm GC}$) are described as the power-law form;
\begin{equation} 
 {\Sigma}_{\rm GC} \propto R^{\alpha},
\end{equation}
we derive the power-law slope ($\alpha$)
for each GCS.
The derived power-law slopes can be compared with
those of stellar halos with 
${\Sigma}_{\rm FS} \propto R^{\alpha}$
so that we can discuss the differences in structural properties
of these two old stellar components. 
The maximum velocities  ($V_{\rm m}$)
and the central velocity dispersion (${\sigma}_{0}$)
in the radial dependences of rotational velocities ($V_{\rm rot}$)
and velocity dispersions ($\sigma$) are derived to
discuss kinematics of the two components.
For convenience,  $V_{\rm m}/{\sigma}_{0}$ is simply referred
to as $V/\sigma$ throughout this paper.

We adopt the slit size of 5 kpc (corresponding to
the effective radius of the simulated GCS) so that we can  estimate
the radial profiles of $V_{\rm rot}$ and $\sigma$
with reasonably small error bars.
Errors in $V_{\rm rot}$ ($\sigma$) 
are assumed to
be equal to $V_{\rm rot}/\sqrt{2(N-1)}$ ($\sigma/\sqrt{2(N-1)}$),
where $N$ is the total number of particles for a given radial bin.
Errors in  $V/\sigma$ are estimated from
the total number of particles at the radii where 
$V_{\rm rot}$ becomes maximum.

The total  number of GCs are $70-450$ (at $z=1$) 
in the models with $z_{\rm trun}=15$
and thus much larger than the observed number (13) of 
old GCs  in the {\rm present} LMC. 
McLaughlin (1999) showed that total number of initial GCs
in a galaxy
can decrease by a factor of 25 within the Hubble time
owing to GC destruction by the combination effect of
galactic tidal fields and internal GC evolution
(e.g., mass loss from massive and evolved stars).
Therefore, it is reasonable
to say that only several percent  of the simulated
GCs can survive to be observed as halo GCs in the LMC.
We thus consider that the above range of GC number 
can be reasonable to be compared with observations.

Table 1 summarizes the parameter values:
Model name (column 1), $M_{\rm t}$ (2),
$\lambda$ (4),  ${\delta}_{\rm i}$ (3), 
$N_{\rm min}$ (5),
$z_{\rm trun}$ (6),
${(\frac{V}{\sigma})}_{\rm FS}$ (7),
and  ${(\frac{V}{\sigma})}_{\rm GC}$ (8).
Here ${(\frac{V}{\sigma})}_{\rm FS}$ 
(${(\frac{V}{\sigma})}_{\rm GC}$)
is the mean
$V/\sigma$ for the three projections
($x$-$y$, $x$-$z$, and $y$-$z$)  for the stellar halo
(GCS).
Our previous simulations of disk galaxy formation
shows that the final spin vectors
of the simulated disks are  similar  to the initial
spin vectors of dark matter halos 
(Bekki \& Chiba 2001). 
We thus consider that the  $x$-$z$ plane of the simulated LMC
corresponds to the disk plane of the LMC
in order to discuss the observed kinematics of the stellar halo
and the GCS.

All the calculations have been carried out on the GRAPE board
(Sugimoto et al. 1990).
Total number of particles used in our simulations is 508686 
and the gravitational softening length is 
0.18 kpc for the LMC models.
The adopted softening length is roughly similar  to the initial
mean separation of the particles in a simulation. 
We used the COSMICS (Cosmological Initial Conditions and
Microwave Anisotropy Codes), which is a package
of fortran programs for generating Gaussian random initial
conditions for nonlinear structure formation simulations
(Bertschinger 1995). 

The present study does not intend to investigate destruction of GCs
and the resultant formation of field stars in the LMC's  halo.
The halo field stars originating from low-mass GCs and initially
unbound or weakly bound star clusters (SCs) may well have
structural and kinematical properties similar to those of the GCS
in the LMC, if the probability of GCs (and SCs) being destroyed
by the combination effects of the LMC's tidal field and internal
GC evolution does not depend on orbits of GCs with respect to
the LMC's center. As described later, the simulated halo and GCS
in the LMC have similar dynamical properties. Thus,  field halo
formation via GC destruction would not change the main results
of the present models.

As described later, the present models do not reproduce 
self-consistently the observed kinematics of halo field  stars
and GCs in the LMC. This is in a striking contrast with
previous simulations (e.g., Santos 2003; Bekki 2005) in which
physical properties of the Galactic GCS can be well reproduced
if truncation of GC formation by reionization is properly modeled.
This is partly because the previous simulations only tried to explain
the Galactic GCS, where remarkable kinematical differences in
halo field stars and GCs are not observed (i.e., they did not
discuss stellar halos and GCSs in less luminous galaxies like
the LMC).

\begin{figure}
\psfig{file=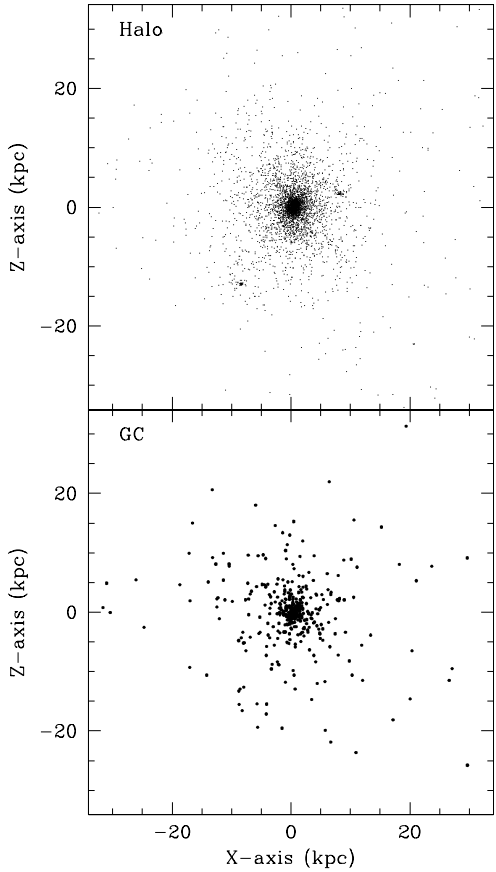,width=6.5cm}
\caption{ 
The final distributions of FS particles 
(upper) and  GC ones (lower)
projected onto the $x$-$z$ plane
in the standard model at  $z=1$.
FS and GC particles form the stellar halo and the GCS,
respectively, in the present study.
The GCS appears to be slightly flattened in comparison
with the stellar halo owing to the smaller number of
GC particles with $|z|>20$ kpc.
The overall distributions are however quite similar with
each other in this model.
}
\label{Figure. 2}
\end{figure}

\begin{figure}
\psfig{file=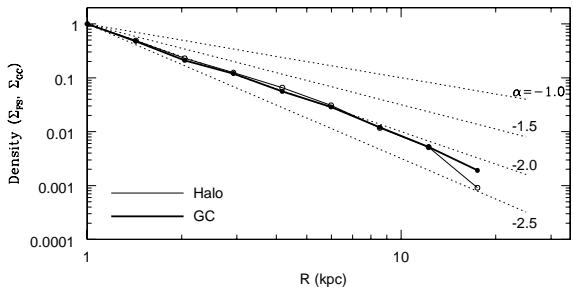,width=7.5cm}
\caption{ 
The final projected radial profiles of FS (thin solid)
and GC (thick solid) particles in the standard model.
For comparison, the profiles normalized to their
central values  are shown.
Both profiles can be well described as power-law ones
with the slopes $\alpha \sim -2.0$ for $R<10$ kpc.
}
\label{Figure. 3}
\end{figure}

\begin{figure}
\psfig{file=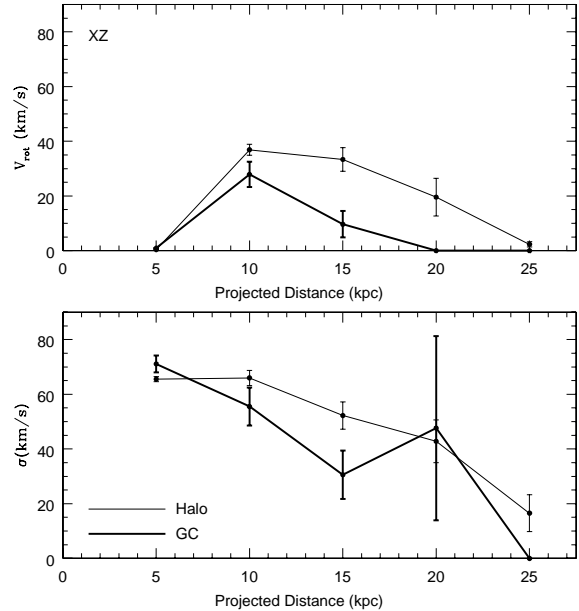,width=7.5cm}
\caption{ 
Radial dependences of rotational velocities $V_{\rm rot}$
(upper) and velocity dispersions $\sigma$ (lower) 
for the stellar halo (thin solid) and the GCS (thick solid)
projected onto the $x$-$z$ plane
in the standard model with the slit size of 5 kpc
for the estimation of $V_{\rm rot}$ and $\sigma$ 
in each bin.
The projected distance here means the distance along
the $x$-axis in the simulation (See Figure 2).
The results are shown for  $0$ kpc $\le x \le$  25 kpc
in the $V_{\rm rot}$ profile
and for all selected particles with $|x| \le 25$ kpc
in the $\sigma$ one.
$V_{\rm rot}=0$ are plotted with no error bars 
for bins with no GC particles (e.g., at $x \approx 25$ kpc).
Although the error
bars are   not small, 
it is clear that both
the stellar halo and the GCS show little rotation.
}
\label{Figure. 4}
\end{figure}

\begin{figure}
\psfig{file=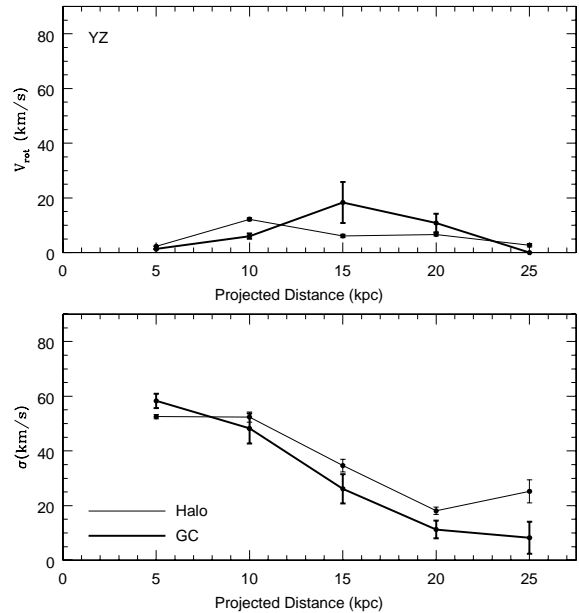,width=7.5cm}
\caption{ 
The same as Figure 4 but for
the halo and the GCS projected onto the $y$-$z$ plane
in the standard model.
}
\label{Figure. 5}
\end{figure}

\section{Results}

\subsection{The standard  model}

Figure 1 shows the time evolution of spatial distributions
of FSs and GCs from $z=30$ to $z=1$ that are formed
within subhalos virialized  before $z=z_{\rm trun}$ (=15).
434 small subhalos are virialized before $z=z_{\rm trun}$ 
and have masses less than 
$3.0 \times 10^7 {\rm M}_{\odot}$ 
at the virialization
and grow via hierarchical merging
with other subhalos with and without FSs and GCs ($z=8.6$).
These smaller subhalos with FSs and GCs 
merge with one another to form bigger subhalos ($z=5.4$),
and finally these bigger halos also merge with one another
($z=3.2$) to form a single halo
till $z=1$. 
FSs and GCs are tidally stripped from the subhalos
during this hierarchical merging 
and consequently dispersed into the halo region 
to form a stellar halo and a GCS.

Figure 2 shows that both the stellar halo (composed of FSs) and 
the GCS (GCs) have similar spherical distributions
in the $x$-$z$ projection,
though the GCSs has  a smaller number of GCs with $|z|>20$ kpc.
There are no significant differences 
in distributions projected onto the $x$-$y$, $x$-$z$, and $y$-$z$
planes between the two components: both components show
flattened distributions in the  three projections
with the major axes aligned with each other. 
These similarities means that the internal structures
between the two components 
are almost identical.
The half-number radius is 5.0 kpc for the stellar halo
and 5.3 kpc for the GCS, which suggests that there are 
no significant differences in dynamical properties between
these two components.

Figure 3 shows that both the stellar halo and the GCS 
have radial  density profiles that can be approximated
by the power-law ones 
with the slopes of $\alpha \sim -2$ at least for $R<20$ kpc.
The apparent lack of flattening in the profile
of the stellar halo  in the
inner part of the simulated LMC  ($R<2$ kpc) is
inconsistent with the best-fit exponential profile 
by Alves (2004),
which shows flattening (or ``core'') of the profile
for $R<1$ degree from the LMC's center.
Although this inconsistency between the simulated
and the observed halos of the LMC has some profound
physical meanings about the formation processes of the
stellar halo (Bekki \& Chiba 2007, in preparation),
we intend to discuss this not in this paper but
in our future papers.

Figures  4 and 5 show that there are no remarkable differences
in the radial dependences of rotational velocities
($V_{\rm rot}$) and velocity dispersions ($\sigma$) 
between the stellar halo and the GCS.
Both components show  overall small $V_{\rm rot}$
($<40$ km s$^{-1}$), radially decreasing $\sigma$ profiles,
and small $V/\sigma$ ($<0.6$).
The estimated
$V/\sigma$ in the $x$-$y$, $x$-$z$, and $y$-$z$ projections
are 0.25$\pm 0.18$, 0.57$\pm0.13$, and 0.21$\pm0.01$, 
respectively,  for the stellar halo,
and 0.47$\pm0.33$, 0.60$\pm0.10$, and 0.31$\pm0.13$, 
respectively,  for the GCS.
The derived small $V/\sigma$ clearly indicates that
both components are dynamically supported by
velocity dispersion rather than by global rotation. 
The results shown in Figures 3, 4, and 5 thus show
that there are no significant differences in dynamical
properties between the two components, though they
originate from different parts of subhalos
virialized before $z=z_{\rm trun}$.

The GCS is formed from
subhalos with different masses and epochs of virialization
so that the spatial distributions of GCs originating
from different subhalos are  different with one another.
For example, GCs  from subhalos with masses more than
$ 10^7 {\rm M}_{\odot}$ at their virialization epochs 
have a half-number radius of 3.5 kpc and thus
a more compact spatial distribution in comparison
with the GCS composed of all GCs.
Such a more compact distribution can be seen
in the stellar halo composed only of
FSs originating from more massive subhalos.
These results imply that stellar components
formed in more massive subhalos
at high $z$  are more likely to
be the inner parts  of galaxies at $z=0$.

\begin{figure}
\psfig{file=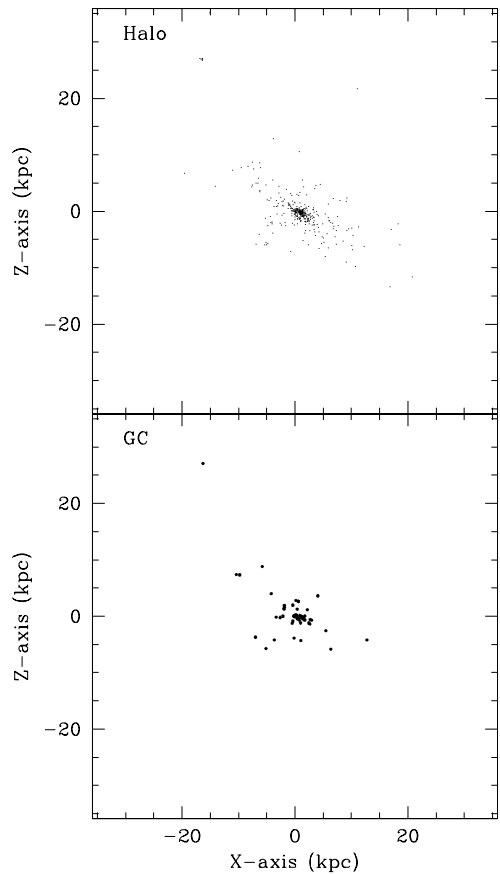,width=6.5cm}
\caption{ 
The same as Figure 2 but for the low-density LMC model.
Note that both the stellar halo and the GCS appear
to be flattened.
}
\label{Figure. 6}
\end{figure}

\begin{figure}
\psfig{file=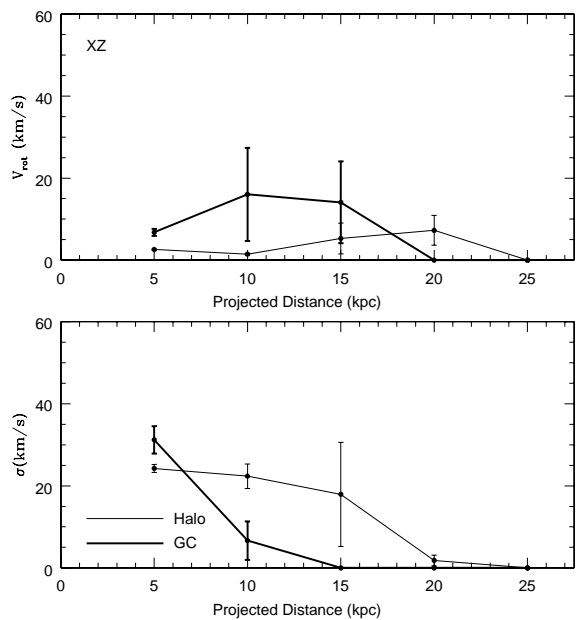,width=7.5cm}
\caption{ 
The same as Figure 4 but for the low-density LMC model.
$V_{\rm rot}=0$ ($\sigma =0$) are plotted with no error bars 
for bins with no GC (or halo)  particles (e.g., $x \approx 20$ kpc
and $x \approx 25$ kpc in the $V_{\rm rot}$ profile of GCs).
}
\label{Figure. 7}
\end{figure}

\begin{figure}
\psfig{file=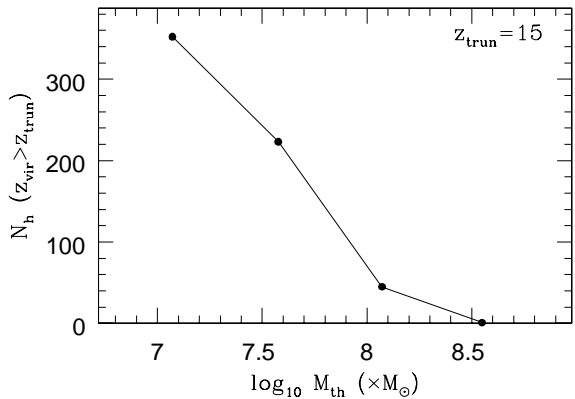,width=7.5cm}
\caption{ 
The dependence of the total number of subhalos  ($N_{\rm h}$)
with GCs,
$z_{\rm vir}>z_{\rm trun}$,  and $M_{\rm h} > M_{\rm th}$
in the standard model with different $M_{\rm th}$
for $z_{\rm trun}=15$.
$M_{\rm h}$ is the  mass of a halo when the halo is identified
as a ``virialized system'' in the  simulation.
Note that $N_{\rm h}$ is very small  
for $M_{\rm th} > 3 \times 10^8 {\rm M}_{\odot}$.
}
\label{Figure. 8}
\end{figure}

\subsection{Parameter dependences}

The dependences of
structural and  kinematical properties
of stellar halos and GCSs  on model parameters 
are summarized as follows.

(i) low-density model: Figure 6 shows that the final 
projected distributions
of the stellar halo and the GCS in the low-density
LMC model with ${\delta}_{0}=0.19$ appears to be
flattened in comparison with the standard model.
We confirm that this flattening can be seen
in the three projections (i.e., $x$-$y$, $x$-$z$, and $y$-$z$)
and thus suggest that the derived  flattening is due to the 
lower initial density of the LMC in this model.
The power-law slope $\alpha$ 
in the projected density distribution is  $-2.5$ both for the stellar
halo and for the GCS, which means that the radial density profiles
are slightly steeper in this model than in the standard one.

(ii) low-density model: 
Figure 7 shows that the two components in the low-density
model have  a small amount of rotation ($V_{\rm rot}<20$ km s$^{-1}$),
which suggests that the two components are dynamically supported
by velocity dispersion. The mean $V/\sigma$  is 0.60 for the stellar
halo and 0.56 for the GCS, which are slightly higher than
those in the standard model (See table 1). Flattened shapes
and small $V/\sigma$ in this model suggest that the two 
components have anisotropic velocity  dispersions.

(iii) high $N_{\rm min}$ model:
the high threshold model with $N_{\rm min} = 1000$ 
has  more compact distributions of the stellar halo 
and the GCS with the half-number radii of the two
equal to 4.5 kpc and 3.4 kpc, respectively.
This is due to the fact that more massive subhalos virialized
before $z_{\rm trun}$ can finally settle in the inner region
of the LMC in this model.
The GCS appears to be more flattened than the stellar halo,
though $V/\sigma$  is not different between the two components.

(iv) low-z truncation model: 
both  the stellar halo and the GCS have
less compact spatial distributions and 
shallower radial density profiles with $\alpha \sim -1.5$ for
the central 10 kpc in the low-z truncation model.  
Both the stellar halo and the GCS have small $V/\sigma$
(0.19 and 0.21, respectively), which means that the two
components are supported by velocity dispersion
more strongly. A large number  of GCs ($\sim 5000$)  
is totally inconsistent with observations, even if
later GC destruction by galactic tidal fields are considered.

(v)  Thus the present models all  show very small $V/\sigma$ ($0.2-0.6$)
of GCSs,
which is much smaller than the observed value of $\sim 2$.
Small $V/\sigma$ can be seen in models
with high spin parameters (e.g.,  $\lambda =0.12$) 
and low initial masses ($M_{\rm t}=2.0 \times 10^{10} {\rm M}_{\odot}$)
and thus suggests that the present {\it dissipationless} models
can not reproduce well the observed rotational kinematics 
of the LMC's GCS.
The inner flattened halo derived only in the low-density model
can be consistent with the observed one by  
Subramaniam (2006)
and thus implies that the LMC could be formed from 
a low-density galaxy-scale fluctuation.

\section{Discussions}

We have shown that (1) there are no significant kinematical
differences in the simulated stellar halos and GCSs 
in the LMC models
and (2) $V/\sigma$ of the GCSs can not be as high
as observed ($\sim 2$) in the GCS of the LMC.
These failures to reproduce the observe kinematical differences
between the two old stellar components (i.e., stellar halo
and GCS) in the LMC imply that
the adopted models for the formation of the two components
at high redshifts can lack some important ingredients of GC formation. 
These failures can be  due to the adopted assumptions
that (i)  all subhalos with different masses and 
different redshifts of virialization can
have {\it both field stars and GCs}
and (ii) the GCS was formed from {\it dissipationless  merging}
of subhalos that had been virialized before reionization 
(i.e., $z_{\rm vir} > z_{\rm trun}$)
and thus had GCs (before their merging leading to the formation
of the LMC at later redshifts).

Observational studies of GCs for galaxies in the Local
Group of galaxies showed that dwarf galaxies fainter
than $M_{\rm V} = -13$ mag appear to have no GCs
(van den Bergh 2000). 
This result means  that (i) there could be 
a possible threshold galaxy 
mass ($10^8 - 10^{9} {\rm M}_{\odot}$) 
above which GC formation is possible 
and (ii) the present dissipationless models discussed so far
did  not
consider this possible threshold mass ($M_{\rm th}$). 
We accordingly investigate how the total number of
subhalos that can be virialized before $z_{\rm trun}$
and thus have GCs ($N_{\rm h}$) 
depend on $M_{\rm th}$ for the 
standard model  with $z_{\rm trun}=15$.
Figure 8 shows that (i) $N_{\rm h}$ is smaller
for the models with larger $M_{\rm th}$ 
and (ii) if  
$M_{\rm th} > 3 \times 10^8 {\rm M}_{\odot}$,
almost no subhalos virialized before  $z_{\rm trun}=15$
can have GCs. The required $M_{\rm th}$ for no halo formation
with GCs before reionization is higher for lower $z_{\rm trun}$.

These results in Figure 8 therefore imply that
if there is a threshold halo mass ($M_{\rm th}$) for GC formation
and if $M_{\rm th} > 3 \times 10^8 {\rm M}_{\odot}$,
the LMC's GCS  can not be formed from hierarchical merging
of subhalos with $z_{\rm vir}>z_{\rm trun}$ and  with GCs.
As shown in the present studies, 
dissipationless merging of subhalos is responsible for
the larger velocity dispersion of old stellar components.
These  results in Figure 8 accordingly imply
that the observed small velocity dispersion
of the LMC's GCS is due to
the fact that the LMC's GCS  was not formed from
dissipationless merging of low-mass  subhalos  
with  $z_{\rm vir}>z_{\rm trun}$.  
Figure 8 thus implies  that old, metal-poor GCs in the LMC
were not formed in low-mass  subhalos that were the building blocks
of the LMC, because the masses of the subhalos  were systematically lower
than $M_{\rm th}$ before reionization.

If the LMC's old GCs do not originate from subhalos
virialized at high redshifts, how were they formed ?
Previous numerical simulations showed
that GCs can be formed during dissipative merging
between the Galaxy and gas-rich dwarfs (Bekki \& Chiba 2002).
We accordingly consider that {\it dissipative merging}
of gas-rich subhalos can trigger the formation of GCs
at the very early epoch of the disk formation of the LMC.
Formation of GCs via dissipative merging
of subgalactic clumps (e.g., gas-rich dwarfs)
in the very early epoch of
galactic disk formation results in disky spatial distributions
and rotational kinematics of GCSs (Bekki \& Chiba 2002).
If old, metal-poor GCs in the LMC
are formed by the above dissipative processes,
they should have slightly younger ages than the Galactic counterparts.

We thus suggest the following  possible scenario
for the origin of the observed kinematical difference
in the stellar halo and the GCS of the LMC: 
(1) the stellar halo was formed from merging of
low-mass subhalos with field stars and with no/few GCs
(i.e., GC-less galaxy building blocks)
and thus shows a large velocity dispersion
and a low $V/\sigma$
and (2) the GCS was formed through  dissipative
merging of gas-rich subhalos and 
gas infall at the very early epoch of the disk formation
and thus shows rotational kinematics.
In the first part (1) of this scenario,
the masses of subhalos virialized
before reionization were well below the
threshold mass ($M_{\rm th}$) for GC formation so  that
the halos could not form GCs.
We thus suggest that  $M_{\rm th}$ can cause
differences in structural and kinematical properties
between stellar halos and old GCs in less luminous galaxies
like the LMC.

Olsen et al. (2004) found that the kinematics of GCSs in 
late-type galaxies in the Sculptor group are 
consistent with rotational kinematics seen in HI components of these galaxies
and suggested that the GCSs were formed in disks rather than in halos.
Beasley et al. (2006) also found a large $V/\sigma \sim 3$ of
the GCS  
in the low-mass dwarf galaxy (VCC 1087) in the Virgo cluster of
galaxies.
These observations imply that the rotational kinematics
seen in the LMC's GCS is not exceptional but can be 
found in GCSs of many less luminous galaxies and thus
that the formation processes of GCSs in these galaxies
can be discussed in terms of the  
proposed scenario above.
We also suggest that there can be  two different formation processes
of old, metal-poor GCs
before and after reionization: one is GC formation
in high-density central regions of subhalos early virialized
before reionization and the other is GC formation
in the very early stage of disk formation. 
Future observational studies on the shapes of the stellar halos
in these galaxies with GCSs having rotational kinematics
will enable us to discuss the origin of 
the possible differences in structures and kinematics
between the stellar halos and the GCSs for these galaxies
in the context of the proposed scenario.

\section{Conclusions}

We numerically investigated structural and  kinematical properties of
the stellar halo and the GCS in
the LMC 
by assuming that the two old components  can be formed
from dissipationless merging of subhalos that
were virialized before reionization and contained both
field stars and GCs.
We particularly discussed
whether or not 
the observed GCS's rotational kinematics
($V/\sigma \sim 2$)  
can be reproduced by the present models.
Our simulations 
with different model parameters
showed that the GCS composed of metal-poor GCs
formed high redshifts ($z > 6$) before reionization 
has little rotation ($V/\sigma \sim 0.4$) and structures
and kinematics similar to those of the stellar halo.
This inconsistency therefore
implies that if  old, metal-poor  GCs in  the LMC
have rotational kinematics,
they are highly unlikely to originate from
the low-mass subhalos
that formed the stellar halo:
the adopted assumption that both the field stars and
the GCs in the LMC  were formed within {\it all low-mass subhalos}
virialized before reionization
is highly likely to be wrong.

We accordingly considered that there could be a threshold 
halo mass ($M_{\rm th}$) above which GCs can be formed 
(i.e., below which only field stars can be formed)
and investigated how the number of subhalos
($N_{\rm h}$) that can be virialized before $z_{\rm trun}$ and have
GCs depends  on $M_{\rm th}$ for $z_{\rm trun}=15$.
We found that if $3 \times 10^8 {\rm M}_{\odot} \le M_{\rm th}$,
the present LMC can not contain GCs formed within subhalos
with the redshifts of their virialization ($z_{\rm vir}$)
larger than that of reionization ($z_{\rm trun}$).
We also  suggested that 
if $z_{\rm trun}$ is lower, the required $M_{\rm th}$ 
(for no GC formation) needs to be higher.
We therefore concluded that if old, metal-poor GCs in the LMC
have rotational kinematics, 
they were not formed in subhalos virialized before reionization.

We suggested a possible
scenario in which the stellar halo was formed
from low-mass subhalos virialized before reionization
and having  no/few  GCs whereas the GCS
was formed at the very early epoch of the disk formation via
dissipative merging of gas-rich subhalos  and gas infall well after 
reionization.
In this scenario, the origin of the observed possible
rotational kinematics of the LMC's GCS 
is closely associated with dissipative gas dynamics
in the disk formation of the LMC.
The LMC's GCs are thus suggested to be slightly younger
than the Galactic counterparts.
It is however unclear how old GCs can be formed 
during dissipative formation of the main body of the LMC.
We thus plan to investigate whether the observed
possible global rotation of the GCS 
in the LMC can be reproduced by our more sophistical
numerical models with gas dynamics and GC formation.

We also suggested that 
the threshold halo mass ($M_{\rm th}$) for GC formation
can cause 
significant  differences in structural, kinematical,
and chemical properties between  stellar halos and GCSs
in less-luminous galaxies like the LMC.
Future observations will extensively investigate
structural and kinematical differences in stellar halos
and GCs 
for galaxies beyond the Local Group
and thus confirm the presence or the absence
of the differences. 
We  plan to investigate dependences of  physical properties 
of stellar halos and GCSs on physical conditions
of their host galaxies at their formation epochs
(e.g., masses and spin parameters) based on
fully self-consistent chemodynamical simulations.

\section*{Acknowledgments}
We are  grateful to the referee for valuable comments,
which contribute to improve the present paper.
KB  acknowledges Masashi Chiba for his useful discussions
on the origin of stellar halos in galaxies. 
KB  acknowledges the financial support of the Australian Research 
Council throughout the course of this work.
The numerical simulations reported here were carried out on GRAPE
systems kindly made available by the Astronomical Data Analysis
Center (ADAC) at National Astronomical Observatory of Japan (NAOJ).

\end{document}